\documentclass{aa}
\usepackage[colorlinks=true,citecolor=blue]{hyperref}
\usepackage{graphicx}
\usepackage{txfonts}
\usepackage{natbib}
\usepackage{deluxetable}

\usepackage{url}
\usepackage{color,hyperref}
\definecolor{linkcolor}{rgb}{0,0.3,0.7}
\hypersetup{colorlinks=true,
	linkcolor=linkcolor, 
	citecolor=linkcolor,
	filecolor=linkcolor, 
	urlcolor=linkcolor}

\begin{document}  

    \title{The optical and near-infrared spectrum of the Crab pulsar with X-shooter}
\titlerunning{The Crab Pulsar spectrum}

\authorrunning{Sollerman et al.}
   \author{J. Sollerman \inst{1} \and 
   J. Selsing \inst{2} \and 
          P. M. Vreeswijk \inst{3,4} \and 
          P. Lundqvist \inst{1} \and 
          A. Nyholm \inst{1} 
          }
          
   \institute{Department of Astronomy, The Oskar Klein Centre, Stockholm University, AlbaNova, 10691 Stockholm, Sweden.\\
            \email{jesper@astro.su.se}
            \and The Cosmic Dawn Center, Niels Bohr Institute, Copenhagen University, 
            Vibenshuset, Lyngbyvej 2,
            DK-2100 Copenhagen, Denmark
    \and Department of Astrophysics/IMAPP, Radboud University, P O Box 9010, NL-6500 GL Nijmegen, The Netherlands
    \and
    Department of Particle Physics and Astrophysics, Weizmann Institute of Science, Rehovot 7610001, Israel
                      }

   \date{Received; accepted}

  \abstract
   {Pulsars are well studied all over the electromagnetic spectrum, and the Crab pulsar may be the most studied object in the sky. 
   Nevertheless, a high-quality optical to near-infrared spectrum of the Crab or any other pulsar has not been published to date.
   }
   {Obtaining a properly flux-calibrated spectrum enables us to measure the spectral index of the pulsar emission, without many of the caveats from previous studies. This was the main aim of this project, but we could also detect absorption and emission features from the pulsar and nebula over an unprecedentedly wide wavelength range.}
   {A spectrum was obtained with the X-shooter spectrograph on the Very Large Telescope. 
   Particular care was given to the flux-calibration of these data.}
   {A high signal-to-noise spectrum of the Crab pulsar was obtained from 300~nm to 2400~nm. The spectral index fitted to this spectrum is flat with $\alpha_\nu=0.16\pm0.07$. 
   For the emission lines we measure a maximum velocity of 1600~km~s$^{-1}$, whereas the absorption lines from the material between us and the pulsar is unresolved at the 
   $\sim$50 km s$^{-1}$ resolution. A number of Diffuse Interstellar Bands and 
   a few near-IR emission lines that have previously not been reported from the Crab are highlighted. }
   {}
             
   \maketitle 

\section{Introduction}

The Crab pulsar is located in the middle of the Crab nebula 
(M1) - the remnant of the supernova that occurred in the year 1054 CE.
This system is among the most- and best-studied astronomical objects in the sky, and {the Crab pulsar itself often serves as the 
object against which other astronomical objects 
as well as instruments are gauged.}

In the optical regime, only a dozen neutron stars have been detected, and optical pulsations are seen only for six 
of the several thousands known pulsars {\citep{sg2002}}. 
When it comes to spectroscopic observations of pulsars in the optical regime, 
most optical pulsars are simply too faint, making such observations only marginally possible 
(for example, \citealp[Vela pulsar,][]{mignani2007}; 
\citealp[PSR B0540-69,][]{serafimovich2004}; 
\citealp[Geminga,][]{martin1998} and 
\citealp[PSR B0656+14,][]{zharikov2007}).

On the other hand, the Crab pulsar is bright at $V$=16.5 and is easily accessible with optical telescopes. 
It is thus the only pulsar for which a decent signal-to-noise spectrum can be obtained {at optical wavelengths.}
Nevertheless, a complete optical to near-infrared spectrum of this canonical object has been lacking.

Following the first optical spectrum {of the Crab pulsar} 
by \citet[][]{oke1969} it took almost 30 years before a modern CCD spectrum was published \cite[][]{nasuti1996}. 
This was followed 
by several investigations that provided data with medium-sized ($2-4$~m) telescopes more than
15 years ago \citep[e.g.,][]{sollerman2000,carraminana2000,beskin2001,romani2001,fordham2002}.
These observations were also followed by a series of theoretical papers aiming to interpret the optical emission
\citep[e.g.,][]{bjornsson2010,massaro2006,oconnor2005,crusiuswätzel2001}.
Yet, while important from a
theoretical perspective, the exact shape of the non-thermal
optical--near-infrared (NIR) spectral-energy-distribution (SED) of 
this isolated neutron star has remained elusive.

While we have previously studied the Crab pulsar SED from the ultraviolet regime
with the Hubble Space Telescope (HST)
\citep[][]{sollerman2000}, to the NIR with VLT/ISAAC \citep[][]{sollerman2003}
and with NAOS/CONICA \citep[][]{sandberg2009}, few
new attempts have been made to improve the observational quality of the Crab pulsar's optical-NIR spectrum. 
The lack of a near-infrared spectrum for this canonical pulsar was surprising already 15 years ago 
\citep[][]{oconnor2005}, 
and 
the NIR spectral energy distribution has instead been patched together using 
photometric data 
\cite[see][]{sollerman2003,sandberg2009} 
from different instruments and occasions.
Despite {decades} of observational and theoretical work, the Crab nebula and its pulsar still holds secrets and surprises, 
as reviewed by 
\citet[][]{hester2008} and by
\citet[][]{0034-4885-77-6-066901}.

In this paper, we present an X-shooter spectrum of the Crab pulsar. 
The main purpose with the X-shooter approach is 
to remedy the above-mentioned situation, by obtaining 
the entire optical-NIR SED with a single instrument.
The very wide wavelength
coverage of the X-shooter spectrograph provides the complete $U$-$K$ spectral energy distribution of the Crab pulsar at the same time. 

A coherent study of the non-thermal emission of an isolated neutron star could 
test the reality of the 
near-IR self-absorption roll-over. 
Synchrotron self-absorption in the NIR regime has been suggested and invoked in pulsar models since the early 1970's, 
but the evidence for this in the Crab pulsar was questioned by
\citet[][]{sandberg2009}. The actual value of the spectral index 
$\alpha_\nu$ (we define the spectral index to be given as the power-law F$_\nu \propto \nu^{+\alpha_\nu}$)
of the pulsar in this region is also 
of importance for understanding the emission mechanism of pulsars. 
\citet[][]{bjornsson2010} showed for example how the value of the spectral index in the optical-NIR regime translates to limits on the location of the emission zone within the pulsar magnetosphere.

All X-shooter data presented in this paper
are publicly available in the ESO archive and the fully reduced flux-calibrated spectrum is uploaded to the WISeREP \citep[][]{wiserep}. 

This paper is organized as follows: 
observations and data reductions are presented in Sect.~\ref{sec:observations}; spectroscopic analysis is provided in Sect.~\ref{sec:spec}. Many of the caveats of pulsar optical spectroscopy are further discussed in Sect.~\ref{sec:caveats}; whereas a discussion of our results 
are given in Sect.~\ref{sec:discussion}. In 
Sect.~\ref{sec:lines} we present some additional science that can be done with these data, using absorption lines from the ISM, Diffuse Interstellar Bands and emission lines from the ejecta filaments. We summarize in
Sect.~\ref{sec:conclusion}.

 \begin{figure}
   \centering
   \includegraphics[width=9cm]{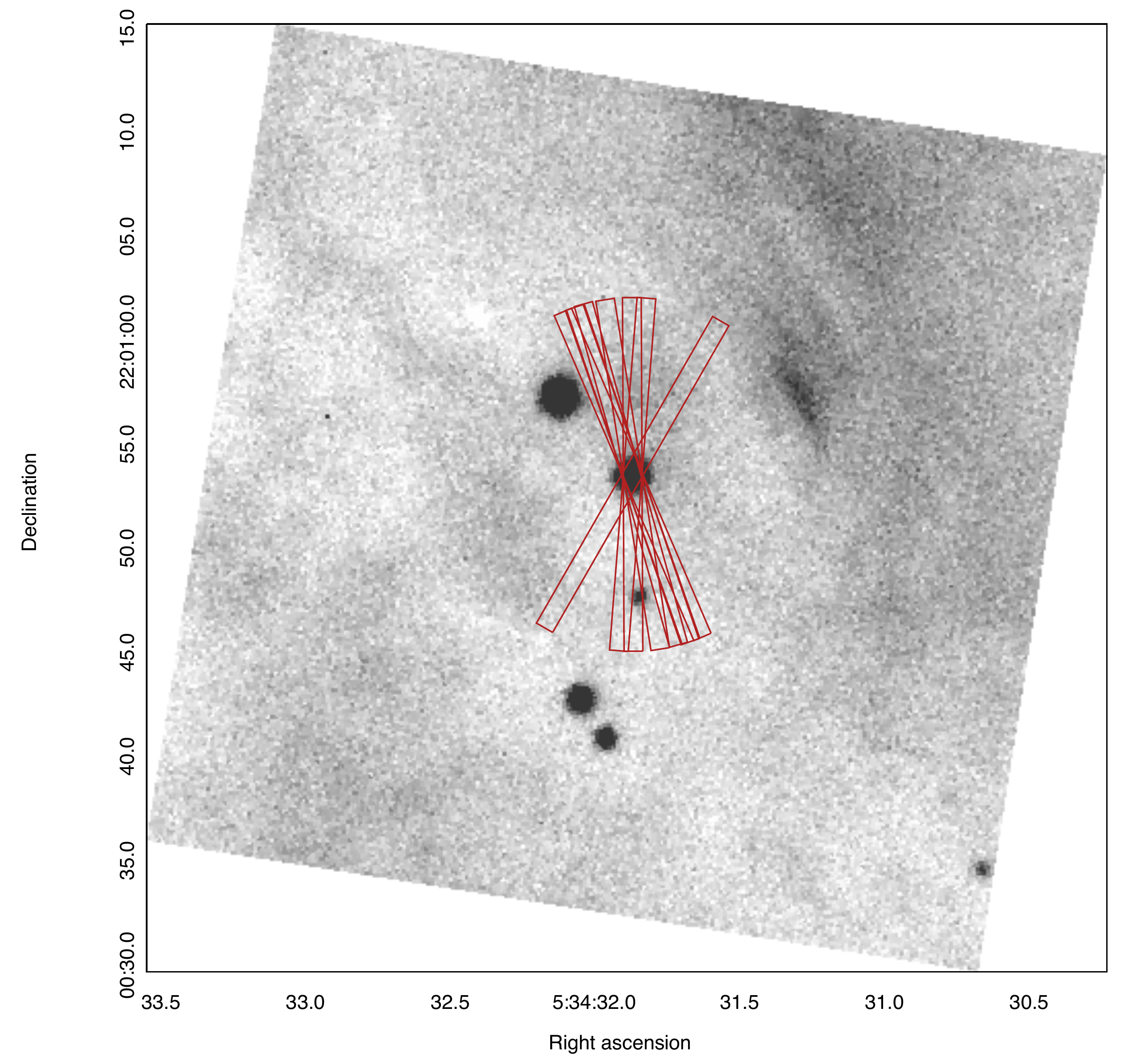}
\caption{
An acquisition image of the Crab pulsar and the neighbourhood from our X-shooter observations. The position angles
provided in Table~\ref{tab:speclog} are overlaid to show where the observations were obtained. The widths of the slits shown are 0\farcs9 and the lengths indicated are 17 arcsec. The latter represents the slit length of 11 arcsec used as well as the 6 arcsec nodding between the ABBA sequences. Coordinates are in J2000.0.
}
\label{Fig0}
   \end{figure}
 
 \begin{figure*}
   \centering
   \includegraphics[width=17cm]{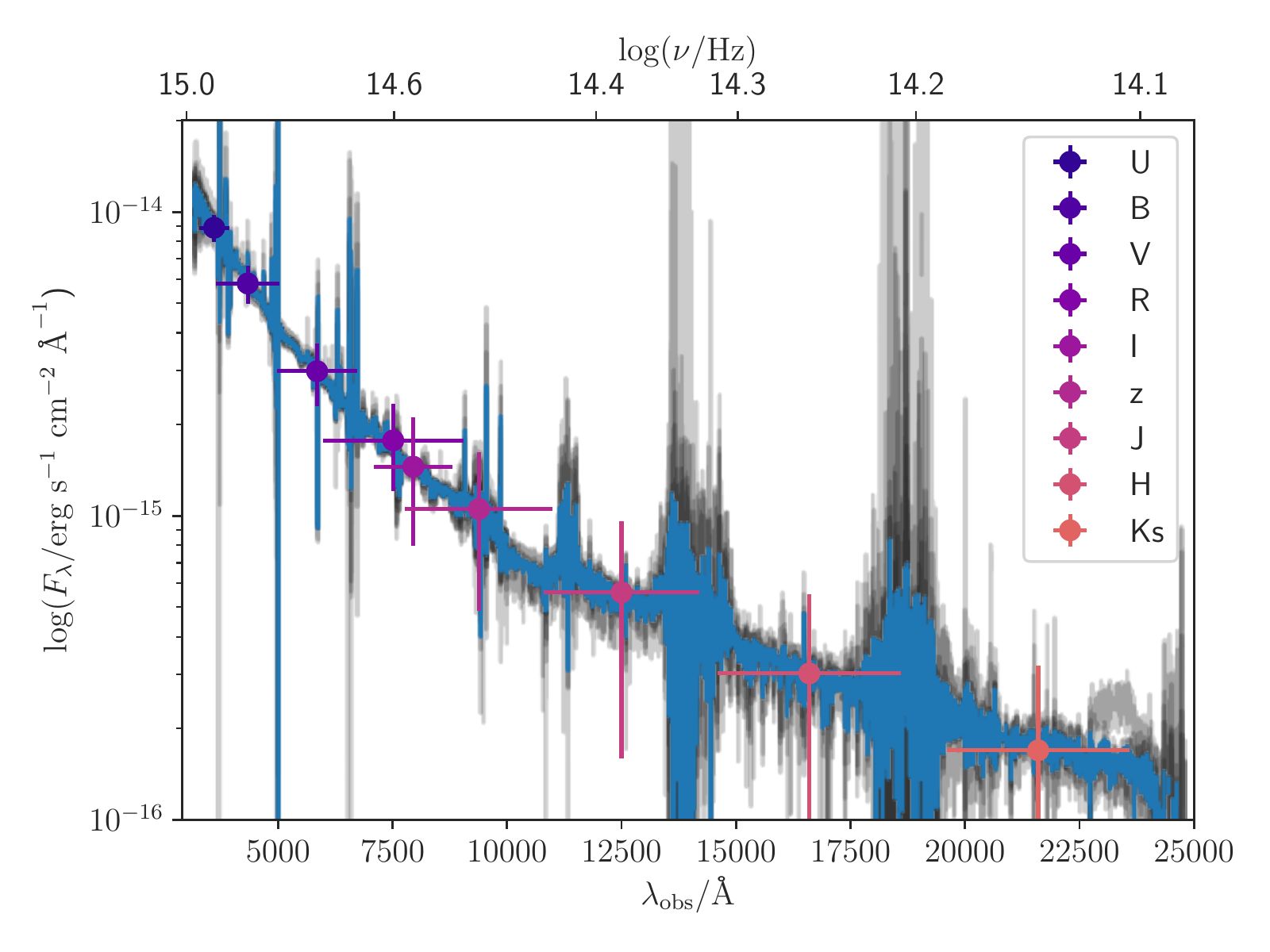}
\caption{
The final fully-reduced and flux-calibrated 
optical-to-NIR X-shooter spectrum of the Crab pulsar. 
The seven individual spectra (Table~\ref{tab:speclog}) are plotted in black, whereas the blue line shows their 
average. As can be seen, all the spectra overlap nicely even though obtained at different occasions. The spectra have been corrected for extinction as discussed in the text. The circles are photometric AB magnitudes from \cite{sandberg2009} - no additional match had to be done.
}
\label{Fig1}
   \end{figure*}

\begin{figure}
   \centering
   \includegraphics[width=9cm]{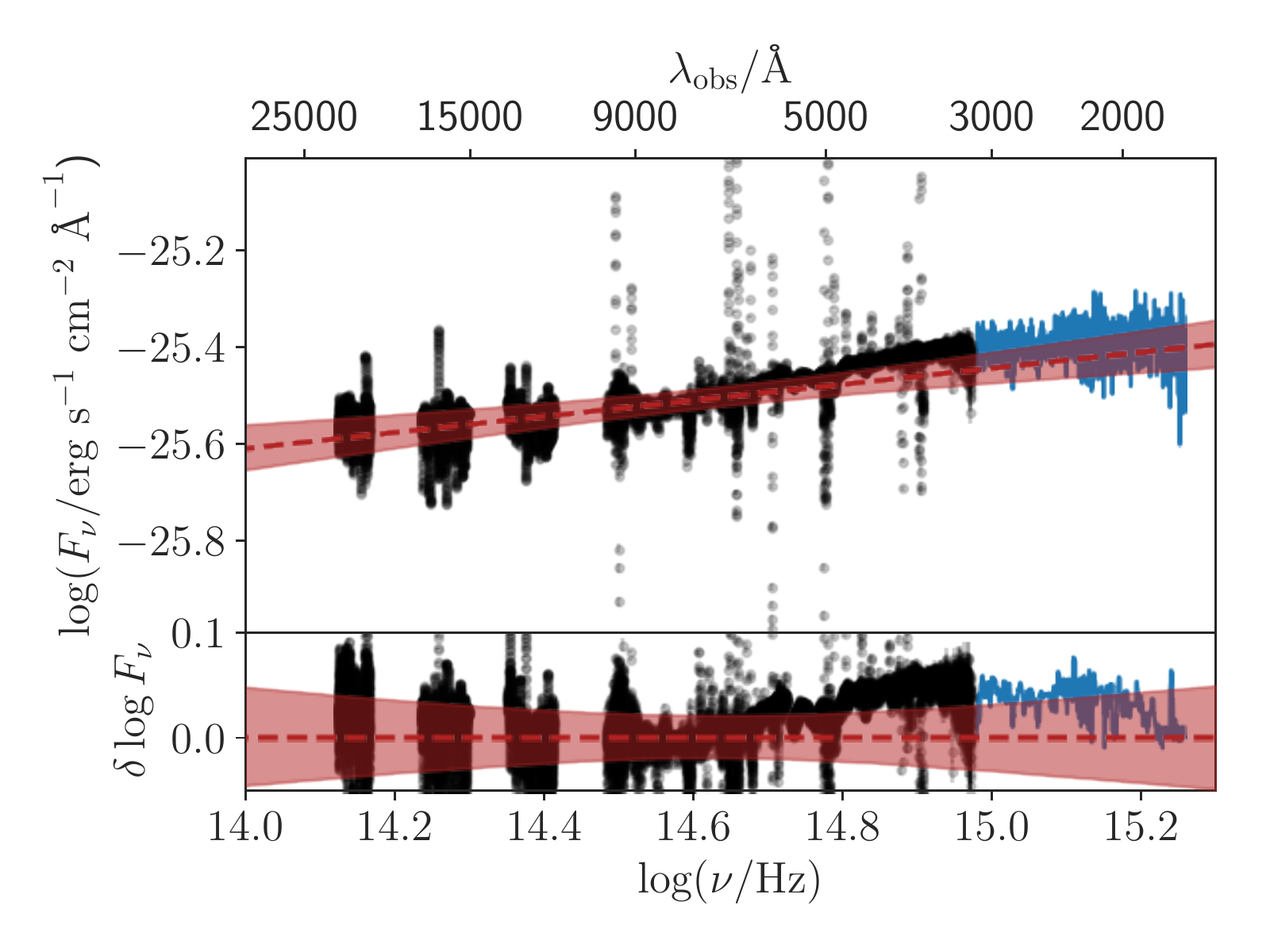}
\caption{
The combined X-Shooter spectrum in log-log space in black together with the best fit power-law in dashed red.
This figure also includes the near-UV HST spectrum from \cite{sollerman2000} in blue. In the fit we have only included the X-shooter spectrum and removed some of the regions where the atmospheric transmission is low, as indicated in the plot. The spectral index is $\alpha_\nu=0.16\pm0.07$ as shown by the dashed line, where the red region encompasses 68\% of the probability mass of the slope uncertainty, normalized at 7000~\AA. The bottom panel zooms in on the residuals between the best fit line and the observed data. 
}
\label{Fig2}
   \end{figure}

\section{Observations and Data Reduction}
\label{sec:observations}

The observations were obtained using the X-shooter echelle spectrograph \citep{vernet2011}
mounted on the European Southern Observatory (ESO) 8.2m Very Large Telescope (VLT), on Cerro Paranal in Chile. This is a 
multi-wavelength medium-resolution spectrograph that simultaneously obtains spectra with three different arms from the ultraviolet to the near-infrared. Each arm is an independent cross dispersed echelle 
with optics and detectors optimized 
for the aimed wavelength regimes.
The total nominal wavelength coverage of the combined (3 arms) spectrum is $3000-24000$~\AA. 
A log of the spectroscopic observations is given in Table~\ref{tab:speclog}, and the field observed in shown in Fig.~\ref{Fig0}.

The strategy for these service mode observations were decided by the aim to obtain the best possible flux-calibrated data to work with. 
This was the main purpose and challenge in this study.

In our first observing run in 2014\footnote{programme 092.D-0260, P.I. Sollerman.} 
we requested the observations to be obtained in dark or grey time under seeing conditions better than $0\farcs8$ and at maximum airmass of 1.7. 
Given the northern position of the Crab pulsar at declination $+22$ degrees, 
the object actually does not rise higher {than 43 degrees above the Paranal Horizon.}
All observations were 
conducted using 420 sec exposures in the 
bluest (UBV) arm and 480 sec in the optical and NIR arms, respectively, for each sub-spectrum. 
We used an ABBA nod-on-slit approach with a nod throw of 6 arcsec. 
A $1\farcs0$ wide slit was used in UVB whereas $0\farcs9$ wide slits were used in the optical and NIR arms.
The purpose of requesting good seeing was to be able to resolve the pulsar from the nebula, 
and the constraint on airmass was to minimize the atmospheric differential refraction, which can otherwise compromise the relative flux-calibration.

For each set of observations, we aligned the slit with the parallactic angle. 
This was done given the airmass of the observations, the lack of a proper Atmospheric Dispersion Corrector (ADC) 
for X-shooter, and the ultimate aim to obtain a carefully flux-calibrated spectrum to determine the spectral index for the pulsar. 
This means that the individual observations presented in Table~\ref{tab:speclog} probe different position angles, 
which are provided in the table, and illustrated in Fig.~\ref{Fig0}. For the pulsar itself this has little effect, but it is relevant 
when investigating 
the neighbouring nebula.

In the first season of Crab pulsar observations (year 2014),
most of the conducted sequences were (marginally) outside specifications of $0\farcs8$ seeing. 
For this reason, we re-applied for observing time for the next batch of observations, 
also with the hope that an operational ADC would then be in place. 
The second run\footnote{programme 096.D-0847, P.I. Nyholm.} 
was requested to be obtained under seeing conditions better than $0\farcs9$ and at maximum airmass 1.7. 
This set of observations was obtained in very favourable conditions. 
We make use of all the available observations in this analysis,
but also 
separate the different datasets due to differences in temporal and seeing conditions.

Using an ABBA nod-on-slit pattern is required for the K band where the sky background is high. 
Each observational sequence was performed in $\sim1$ hour blocks, 
requiring some overheads.
Nodding fully out of the nebula could improve the background subtraction, but would have been too expensive in terms of overheads. The complete observing programme was 4 hours on both occasions. This included extra time 
for the flux-calibration, since we decided to observe an additional flux-calibration 
spectrophotometric standardstar, GD 71, one of very few flux standards with measurements available all the way out into the K band 
\cite[e.g.,][]{vernet2008,fluxxshooter}.
This standard star was positioned close to the Crab pulsar on the sky, 
and was observed with an identical ABBA sequence, i.e. we performed this special observation using the same exposure-time and slit width as for the pulsar, so that exact comparisons of the flux-calibration accuracy could be done.  
The standard stars provided by ESO are instead obtained with a wider (5 arcsec) slit. Given that the absolute calibration of our standard star - as done with the ESO flux standards - was excellent (see Appendix), as was the consistency of all Crab pulsar spectra with previous multi-band photometry (Fig.~\ref{Fig1}) we are 
confident in the absolute calibration of the resulting data.

\tabcolsep=0.11cm
\begin{table*}
\caption[]{Log of spectroscopic observations.} 
\centering
     \scalebox{0.85}{
     \begin{tabular}{llllllcrl}
        \hline
        \noalign{\smallskip}
        Object & UT Date & UT Time & Exp. Time$^{\rm a}$ &  Airmass & Seeing & Slit Width$^{\rm a}$ & Pos. Angle & Program ID \\
        & (year-mm-dd) & (hh.mm.ss)& (min) & & (arcsec)&(arcsec)& (deg)&\\
        \noalign{\smallskip}
        \hline
        \noalign{\smallskip}
        Crab & 2014-02-05 & 01:07:30 & $4\times7/8/8$ & 1.46-1.47 & 1.0-1.2 & 1.0/0.9/0.9 & $+$175.5 & 092-D-0260 \\
        
        Crab & 2014-02-20$^{\rm b}$ & 01:26:41 & $4\times7/8/8$ & 1.54-1.67 & 1.0-1.2 & 1.0/0.9/0.9 & $-$160.7 & 092-D-0260 \\
        
        Crab & 2014-02-21 & 01:36:13 & $4\times7/8/8$ & 1.58-1.74 & 1.0-1.2 & 1.0/0.9/0.9 & $-$156.8 & 092-D-0260 \\
        
        Crab & 2014-02-22 & 01:16:32 & $4\times7/8/8$ & 1.53-1.67 & 1.0-1.3 & 1.0/0.9/0.9 & $-$161.1 & 092-D-0260 \\
        
        Crab & 2015-12-08 & 05:43:49 & $4\times7/8/8$ & 1.47-1.55 &   0.6   & 1.0/0.9/0.9 & $-$171.0 & 096-D-0847 \\
        
        Crab & 2015-12-13 & 03:12:30 & $4\times7/8/8$ & 1.64-1.52 &   0.7   & 1.0/0.9/0.9 & $+$150.1 &096-D-0847 \\
        
        GD71 & 2015-12-13 & 04:10:08 & $4\times7/8/8$ & 1.37-1.32 &   0.7   & 1.0/0.9/0.9 & $+$158.8 & 096-D-0847 \\
        
        Crab & 2015-12-13 & 04:59:43 & $4\times7/8/8$ & 1.46-1.49 &   0.7   & 1.0/0.9/0.9 & $-$179.7 & 096-D-0847 \\
        
        Crab & 2015-12-13 & 05:46:18 & $4\times7/8/8$ & 1.51-1.62 & 0.7-0.8 & 1.0/0.9/0.9 & $-$164.4 &096-D-0847 \\
        \noalign{\smallskip}
        \hline
     \end{tabular} 
  }
 \tablefoot{
        \tablefoottext{a}{The three values listed correspond to those of the respective UVB/VIS/NIR arms.}
\tablefoottext{b} The data from this OB are not used. The pulsar was only observed in the first A position, and was not present in the other BBA exposures.
 } 
\label{tab:speclog}
\end{table*}

\subsection{Spectroscopy}

The spectra of the pulsar, the flux standards, and telluric standard stars
were reduced to bias-subtracted, flat-fielded, rectified, order-combined, flux- and wavelength-calibrated images for all three arms (UVB, VIS, and NIR) using the Reflex package and version 3.2 of the X-shooter pipeline \citep[][]{modigliani2010}. The standard stars and the telluric standards were both extracted by the ESO pipeline, in which an aperture the size of the nodding window, centered on the 2D-spectrum is used for standard extraction. For the extraction of the object spectra, we followed the data reduction steps outlined in \cite{selsing2019}, including the recalibration of both the wavelength solution and the application of a slit-loss correction. The wavelength solution is refined by cross-correlating the sky spectrum with a synthetic sky spectrum generated using the ESO Skycalc tool \citep{Noll2012, Jones2013}. After refinement of the wavelength solution, the spectra are corrected for barycentric motion. The slit-loss correction is estimated using the average DIMM seeing ($V$-band seeing at zenith, measured on the mountain, \citealt{Sarazin1990}) across the spectral integration window for each of the observations. The DIMM seeing is further corrected for the average airmass of the observation.
Using the wavelength dependence of seeing \citep{Boyd1978}, we can evaluate the effective seeing, at each wavelength bin. We synthetize a 2D-moffat profile, using the seeing at each wavelength, and the size of the spectroscopic slit is integrated over, yielding the throughput at each wavelength sampling of the spectrum. This can then be used to correct for the slit losses. The nightly ESO flux standard was observed with a wider slit (5 arcsec) than used for the object spectra 
(Table\ref{tab:speclog}), therefore slit losses do not affect the flux standard observations. The telluric standard star observations were carried out using the same observational setup as the science observations, however because the absolute scaling of these are not important, we did not correct these for slit losses. 

The telluric transmission spectra are found by fitting the telluric standard stars with a model atmosphere, using MolecFit \citep{Smette2015}. The science spectra are then corrected for the telluric transmission, by dividing by the atmospheric transmission for both the flux and error spectra. In this way, the error in the regions most affected by telluric absorption is increased, which ensures that a weighted fitting procedure correctly down-weight such
regions. 

To allow future assessments of the systematics of the reduction, for example the accuracy of telluric corrections and merging of spectral orders, 
{we also make the flux-calibrated version of the spectrum from the spectrophotometric standard star GD71 available at WISeREP \citep{wiserep}, together with the final reduction of the pulsar spectra.}

The resolution of each of the individual observations are found by fitting a series of unresolved telluric absorption lines with Voigt profiles. In more than half of the observations, the seeing is better than the slit widths and thus the actual spectral resolution is better than the nominal one. The nominal spectral resolutions are 55, 34 and 50~km~s$^{-1}$ in the three arms, respectively. In the best case, the fifth spectrum taken
(Table~\ref{tab:speclog}), the seeing is only 0\farcs6, 
for which we derive an effective delivered resolution of 
19~km~s$^{-1}$ 
in the VIS arm and 
45~km~s$^{-1}$ 
in the NIR arm. There are no telluric lines in the UVB arm, so assuming that the average relative improvement in the resolution is the same would give 
38~km~s$^{-1}$ 
in the UVB arm.

\section{Spectral analysis}
\label{sec:spec}

The individual spectra 
(Table~\ref{tab:speclog})
were reduced independently and exhibit small differences in the absolute flux scaling, with excellent agreement in the spectral slope (Fig.~\ref{Fig1}). 
To assess the agreement in the absolute flux level, we measure the flux density in a small region ($\sim50$~\AA~wide) around 7000~\AA. 
We measure $2.00\pm0.158\times~10^{-15}~\mathrm{erg}~\mathrm{s}^{-1}~\mathrm{cm}^{-2}$~{\AA}$^{-1}$, where the error is the standard deviation of the flux densites. We scale all spectra to the single spectrum that minimizes the difference between the observed spectra and the measured photometry from \citet{sandberg2009}, see Fig.~\ref{Fig1}. 
This error is also a measure of the precision of the absolute flux calibration, 
which suggests that the absolute scaling of the flux calibration is consistent to within $8 \%$. This is in agreement with the accuracy reported by 
ESO\footnote{\url{http://www.eso.org/observing/dfo/quality/PHOENIX/XSHOOTER/processing.html}}. 
This exercise shows that the variations due to seeing, airmass and use of different spectrophotometric standard stars were small, after the corrections described in the previous section are employed. 

After the spectra have been calibrated and rescaled, we combine them using a simple mean with the errors added in quadrature. We do not use the inverse variance weighting scheme, because due to the brightness of the source, the error is dominated by Poisson noise and thus the inverse variance scheme will be biased. 

To correct for the Milky Way extinction, we used a color excess of $E(B-V)_{MW}= 0.52$~mag \citep{sollerman2000}, and a standard ($R_V=3.1$) reddening law from \cite{fitzpatrick2007}. We use a Python implementation of the reddening laws\footnote{\url{https://github.com/karllark/dust_extinction}}. 

In Fig.~\ref{Fig1}, we compare the individual spectra, overlaid with the combined one, with previously published photometry in this wavelength range from \citet{sandberg2009}. We did {not further} scale the spectra, nor the photometry, and the two types of measurement are consistent within the errors across the entire wavelength coverage. This gives 
credibility to both the absolute flux scaling and to the shape across the spectral coverage of X-shooter. There are pros and cons with all these data, as discussed in Sect.~\ref{sec:caveats}, but the overall match is excellent.

In Fig.~\ref{Fig2} we plot the data in log~F$_\nu$ versus log~$\nu$ as is commonly done in the pulsar literature. In this figure we have also included the near-UV HST data from \cite{sollerman2000}. The details of these data can be found in that paper, but we mention that we have now also uploaded this spectrum to WISeREP. The agreement between the slope and absolute flux scaling between the X-shooter spectra and the near-UV HST data additionally supports the validity of the absolute flux calibration. 

We estimate the spectral index of the Crab pulsar, by fitting a power law to the full spectral range of X-shooter. In this part, care was taken to assess which parts of the spectra were reliable. All fits used the propagated errors from the reduction pipeline as weights, but we furthermore excluded a few wavelength ranges, like the strong telluric bands between the $JHK$ bands in the NIR, and also the first order in the $J$ band where the merging with the VIS arm was poor.

We fit for the spectral index in each individual spectrum
(Table~\ref{tab:speclog}). The best fit parameters of the power law for each individual spectrum and the associated errors on the best fit values are 
found by sampling the posterior probability distributions of the parameters,
assuming flat priors on all parameters.  We use
\texttt{LMFIT} \citep{lmfit} for the fitting implementation, which runs
\texttt{emcee} \citep{emcee} to carry out the sampling of the posterior probability distribution. We initiate 10 samplers, each
sampling for 5000 steps, but discard the first 500 steps as a burn-in phase
of the MCMC chains.  We use the median of the marginalized posterior
probability distribution as the best-fit values, and the 16th and 84th
percentiles as the uncertainties. 

The individual slopes are then averaged, with the statistical uncertainties from the fits propagated. 
The best fit power law provides a spectral index of $\alpha_\nu = 0.16 \pm 0.07$.  The formal error on the  individual linear fits is merely 0.001. 
To properly assess the systematics, we have fitted power-laws to all the seven extracted spectra, and the RMS of these provided spectral indices is 0.065. We add in quadrature the statistical errors from the fits and the standard deviation between the individual best fit slopes to propagate the error on the derived spectral index, and the error on the slope is here thus dominated by the variation between the individual spectra.

\section{Caveats}
\label{sec:caveats}

Obtaining the correct and precise spectral index for the Crab pulsar is 
a challenge, even if we now have a spectrum of this 16.5 magnitude source with an 8 m telescope. 

The ideal pulsar spectrum should be multi-wavelength, phase-resolved, 
high-resolution, background subtracted, simultaneous and properly 
flux calibrated. 
None of the efforts performed so far achieve all of these. We briefly discuss some of these caveats below.

\subsection{Simultaneous and multi-wavelength}

Most of the spectra in the literature cover only a fairly small spectral window in the optical, typically 4000 -- 8000~\AA. 
This range is rather limited when trying to assess the full spectral energy distribution from the ultraviolet to the infrared, 
or for many pulsars even extending from gamma-rays to radio \citep[e.g.,][]{kuiper2001}. 

Moreover, as limited spectral regions are patched together to construct the entire SED of the pulsar, there is the risk that observations obtained at different occasions are not easily matched. There is both the observational challenge in merging datasets from different instrumentation and observing conditions (e.g., seeing, airmass). For the Crab pulsar there is also the notion that the object itself may be intrinsically varying, both on the long time scales of secular evolution \citep{sandberg2009} 
and on the variable background in terms of wisps and knots \citep[][]{hester1995,melatos2005,
rudy2015}.

Our X-shooter spectrum is a 
step in the right direction, affording a wide wavelength range at the same occasions. However, with the present spatial resolution we can not guarantee that we are not probing the inner knot, where the importance is larger towards the reddest wavelengths \citep{sandberg2009}. 

What we can see from Fig.~\ref{Fig1}, however, is that the mis-match is not severe, given that all seven of our X-shooter spectra overlap 
nicely with previous photometric measurements.

\subsection{Phase-resolved and high-resolution for background subtracted}

With X-shooter, we can only aim at the integrated spectrum of the pulsar. Other kind of instrumentation is needed to phase-resolve the spectrum into the different components, as done by e.g., 
\citet[][]{fordham2002,sollerman2000}. 

This may not be a problem per se, but phase-resolved data may allow better background subtraction. Our data, even if obtained at sub-arcsecond seeing, may well be influenced by the nearby knot and other nebular structures. In fact, in nodding on the slit we are always within the nebula (which is very large on the sky at $6\arcmin\times6\arcmin$). Perfect background subtraction is therefore difficult in our data, even if the pulsar itself clearly dominates the signal. 

\subsection{Flux-calibration}
In the set-up and execution of this project we made sure that a special spectrophotometric standard star GD71 
was co-observed with the pulsar, in order to allow assessing how good the overall relative flux-calibration is. High/medium-resolution spectrographs like X-shooter are seldom designed to perform accurate flux-calibration, so this 
is also partly an investigation into how well this can be achieved, following previous such efforts \cite[see e.g.,][]{fluxxshooter,pita2014}.
We present a short investigation of the flux-calibration accuracy of GD 71 in the Appendix.
We have done as good as we can with flux-calibration of the spectrum - given the caveats listed above. The end result is a good signal-to-noise spectrum all across the UV-NIR wavelengths at medium resolution. However, there will always be room for improvement, and the deviations that we highlight in the bottom panel of Fig.~\ref{Fig2} clearly includes a number of effects, including uncertainties in order overlap, imperfections from telluric corrections, absorption lines and DIBs as discussed in the next section as well as the caveats discussed above. This must be appreciated if discussing them in terms of pulsar physics.

\section{Discussion}
\label{sec:discussion}

The theoretical discussion in \cite{bjornsson2010} highlighted that
the main question regarding $\alpha _\nu $ is whether or not it is consistent with a value of 1/3. This is the highest possible value for optically thin incoherent synchrotron radiation as the emission mechanism. They related this to both the electron distribution cut-off and to the 
the magnetic field in the emission region.  
We here estimate 
$\alpha _\nu = 0.16 \pm 0.07 $ and this therefore remains completely consistent with a simple synchrotron radiation process. We see no strong evidence for a change in the slope at NIR wavelengths, as was previously discussed in the context of synchrotron self-absorption \citep{oconnor2005}. In this sense, we confirm previous studies of the optical-NIR spectral index of the Crab pulsar, but with new confidence given the simultaneous and systematic multiwavelength approach.
{There is also room for improvement as mention in the previous section. Overall, the single power-law (PL) is not a prefect fit to the X-shooter spectrum, as emphasized in the residual panel 
of Fig.~\ref{Fig2}. A fit allowing a broken PL would in fact be formally preferred (AIC) but is also not convincing. 
The best fit two-component PL would have $\alpha_1 = 0.03 \pm 0.02$ and
$\alpha_2 = 0.21 \pm 0.04$ with the break at
log($\nu_{\rm{break}}$) = $14.5\pm 0.1$.
Here, the spectral index in the redder region ($\alpha_1$) is flatter, which is the opposite to an infrared roll-over as suggested by \cite{oconnor2005}. We note in passing that their synchrotron self-absorption scenario for the Crab pulsar also implied a low NIR flux for the pulsar in SNR 0540, something that later observations did not confirm 
\citep{mignani2012}. 
The somewhat steeper slope in the blue would instead overshoot the data in the NUV regime, instead of underpredicting it as for the single PL.
For other optical pulsars, there have been discussions on broken power-law spectra, but the quality of the data \cite[see e.g.,][their fig. 4]{mignani2007} and the interpretation is unclear, so we do not discuss this further here.}

\section{Additional science}
\label{sec:lines}

Flux calibration of the pulsar spectrum was the overall aim of this observational campaign. In this section we mention a few other results 
that can be deduced from the data, although we emphasize that the observations were not set up to optimize this kind of science.
We briefly mention searching for absorption lines associated to the pulsar, interstellar medium or the supernova ejecta. 
We also detect a large number of emission lines from the ejecta filaments close to the pulsar.
In particular, some of the NIR emission lines that we detect with the short slit centered on the pulsar have, to our knowledge, not been reported before from the Crab nebula. All data are available in the ESO archive if future projects want to make further use of these observations. 

\subsection{Cyclotron lines and Diffuse Interstellar Bands}

The full coverage high signal-to-noise spectrum allows us to {search, for example, after potential cyclotron lines}. The occurrence of such lines in pulsar spectra were proposed by e.g.,
\citet[][]{romani2000} 
and putatively detected in the first modern 
high-quality spectrum of the Crab pulsar \citep[][]{nasuti1996}. However,
subsequent observations did not detect these features \citep[][]{sollerman2000}. \citet[][]{beskin2001} also reported non-detections of the particular feature, 
but cautioned - given similar claims for the Geminga - that the Crab pulsar cyclotron lines could be time-dependent.

We searched our spectra for any outstanding broad absorption or emission line. 
Again, no support for the putative cyclotron lines were detected.
In particular, the region around 6000~\AA, where \citet[][]{nasuti1996} tentatively detected a broad feature has high signal-to-noise but no sign of a
broad feature in our spectrum.

On the other hand, we do detect a number of diffuse interstellar bands (DIBs). 
Again, the combined spectrum is not ideal for a deep and systematic search for such features, given the contamination of filament emission-line background residuals, but on inspection we could indeed detect most of the stronger DIBs 
\citep{herbig1995,sollerman2005}. 
We see for example a broad feature at 4430~\AA~with an equivalent width (EW) of 1.1~\AA~and a full width half maximum (FWHM) of 20~\AA~as measured using {\tt IRAF splot}. This is one of the strongest DIBs in the list of \cite{herbig1995}. Since carefully searching for DIBs  is somewhat out of the scope for this paper, 
we simply went through the list of DIBs from \citet{sollerman2005},
and tabulate the ones we detected in Table~\ref{tabledibs}.
There is room for improvement in identifying more DIBs in these spectra.
It could also be of interest to correlate the strengths of the DIBs in this line-of-sight since the extinction is quite high and well characterized \citep{sollerman2000}.

\subsection{Emission lines from the Nebula}

Our pulsar exposures also include
emission from the nebular filaments. 
In the UVB-VIS arms we see a multitude of lines with complex kinematics. 
Of course, emission line studies of the Crab nebula have been performed for decades, and the central region probed with our short slit is not where the very strongest emission in the Crab nebula resides.
We can measure for example a maximum velocity of 
[\ion{O}{II}]~$\lambda\lambda 3727,3729$ of  
1287 km~s$^{-1}$ (blueshifted) and 1046 km~s$^{-1}$ (redshifted). 
Similarly, for the strong [\ion{O}{III}] $\lambda 4959$
we get 1402 and 1611 km~s$^{-1}$, respectively, and H$\alpha$ also gives two shells with 
$\pm1400$~km~s$^{-1}$.  

This is all in accordance with the ejecta velocities previously reported in the literature for the Crab nebula. 
Apart from hydrogen lines we detect for example; 
\ion{He}{I} $\lambda\lambda 3889, 4026, 4471, 5876, 6678, 7065$, 
\ion{He}{II} $\lambda 4686$, 
[\ion{C}{I}] $\lambda\lambda 9824,9850$, 
[\ion{N}{II}] $\lambda\lambda 6548,6583$, 
[\ion{Ni}{II}] $\lambda 7377$, 
[\ion{Ne}{III}] $\lambda\lambda 3869,3968$, 
[\ion{Ar}{III}] $\lambda 7136$, 
[\ion{S}{II}] $\lambda\lambda 6717,6731$, 
[\ion{S}{III}] $\lambda\lambda 9069,9531$,
[\ion{O}{I}] $\lambda\lambda 6300,6364$, 
\ion{O}{I} $\lambda 7772$, 
[\ion{O}{II}] $\lambda\lambda 3727,3729$ and $\lambda\lambda 7319,7320,7330$, as well as  
[\ion{O}{III}] $\lambda\lambda 4363,4959,5007$.

\begin{figure}
   \centering
   \includegraphics[width=9.cm]{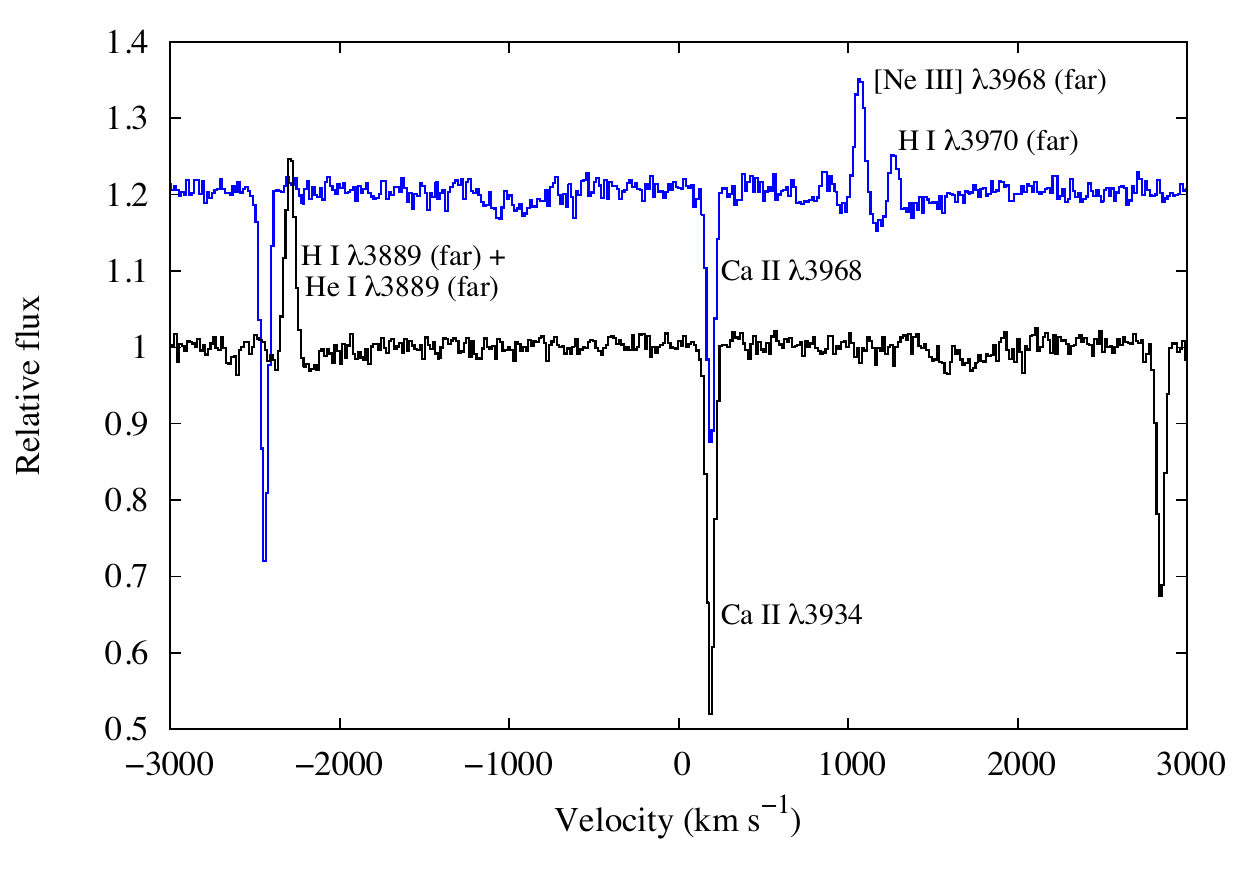}
\caption{
Spectra (in velocity units) around \ion{Ca}{II}~$\lambda\lambda$3934,3968. The spectrum in black centers on \ion{Ca}{II}~$\lambda$3934, whereas the spectrum in blue is centered on \ion{Ca}{II}~$\lambda$3968 and has been shifted {$+0.2$ in relative} flux units for clarity. The narrow, albeit unresolved, absorption close to zero velocity is interstellar absorption. No \ion{Ca}{II} absorption originating in the Crab Nebula can be identified. Emission lines from the nebula (all from the nebula's far side) have been marked.    
}
\label{CaII}
   \end{figure}

Since the optical emission lines have been well studied, 
we focus here instead on the NIR part of the spectrum.
Surprisingly few NIR
spectroscopic studies of the Crab nebula exist,
and for many years the most comprehensive study
\citep{graham1990}
barely detected the strongest
[Fe II], and molecular H$_{2}$ lines.
More recently,
\cite{richardson2013} discussed the NIR spectroscopic observations (K band) of \cite{loh2012}
which mention only observations of \ion{H}{I} and H$_2$ emission from some strong knots.
There has 
been revived interest in the infrared community on the Crab nebula, also in trying to understand the dust mass from modeling 
\cite[e.g.,][]{temim2013,owen2015}.

Clearly, there are a number of disadvantages with our set-up, which was not  optimized to search for - or measure - emission lines. First of all, the X-shooter slit is only 11 arcsec long, and thus covers a very small part of the nebula - the very inner part that is mostly synchrotron dominated (Fig.~\ref{Fig0}). Additionally, the many exposures covered different regions due to the fixation at parallactic angle. Finally, the nodding was done within the nebula, so proper sky subtraction would also remove any uniform nebular emission lines.


Nevertheless, given the apparent lack of Crab nebula NIR emission lines reported in the literature, we have tabulated a number of the detected lines in Table~\ref{tableNIR}. We did this by searching through the 2D image for one of the obtained OBs (no. 6), and tabulated the most conspicuous lines. The spectra are available for further studies. Apart from [\ion{Fe}{II}] 1.64~$\mu$m which was detected by \cite{graham1990} and He~I~2.1~$\mu$m which was seen by \cite{loh2012}, we also detect for example 
[\ion{S}{II}], [\ion{P}{II}] as well as a number of additional iron lines. Phosphorus was also recently detected in Cas~A \citep{koo2013}, where they interpreted the high abundance as evidence for stellar nucleosynthesis. We identify all the six lines that Koo et al. found in Cas A also in our Crab spectra. 

It is obvious that a systematic approach targeting some of the brighter knots in the Crab Nebula 
with X-shooter would deliver a very rich emission line spectrum for nebular diagnostics.

\subsection{Absorption lines}

Finally, we also
searched for
absorption lines from the ISM and the Nebula itself, using the pulsar as a background source. 
This has been suggested 
as a way of probing
the possible fast shell around the Nebula 
\citep[c.f.,][]{lundqvist1986,sollerman2000,tziamtzis2009b}. 

We scanned through both the combined pulsar spectrum,
as well as the spectrum from OB6, 
in order to search for
absorption lines. As displayed in Fig.~\ref{CaII}, 
we clearly detect, for example, 
\ion{Ca}{II} $\lambda\lambda$3934,3968 
in absorption at about rest wavelength. The lines are
very narrow, and 
unresolved at the resolution 
of X-shooter.
\cite{Tappe2004} used VLT/UVES to resolve
several ISM absorption components, and derived 
interstellar column densities, $N_{\rm col}$ (in 
units of cm$^{-2}$), of \ion{Ca}{II}, \ion{Na}{I} 
and \ion{K}{I} to be 
log($N_{\rm col}$) = $13.03\pm0.11, 13.36\pm0.46$ and $11.84\pm0.03$, respectively. Those numbers can be
used, in combination with our measured equivalent
widths of several DIBs, to relate extinction,
DIB strengths and interstellar absorption lines.

Figure~\ref{CaII} shows no hint 
of \ion{Ca}{II}~$\lambda 3934$ absorption in the 
velocity interval $-2000$~km~s$^{-1} < v < 0$~km~s$^{-1}$, and particularly 
none at $\sim -1700$~km~s$^{-1}$, as was evidenced
for \ion{C}{IV} $\lambda\lambda 1548,1551$ \citep{sollerman2000}. 
Our deep X-shooter spectra
can thus rule out similar \ion{Ca}{II} $\lambda 3934$
absorption with much greater confidence than the
data discussed by \citet{lundqvist2012}, which is
obvious from a comparison between Fig.~\ref{CaII}, 
and their fig.~6. Any \ion{Ca}{}
present in the near side of the Crab Nebula must be
ionized to \ion{Ca}{III}, or higher.

\section{Conclusions}
\label{sec:conclusion}

To summarize, we have used the X-shooter three-arm echelle spectrograph on VLT to obtain a UV-NIR spectrum of the Crab pulsar. Summing seven individual but mutually consistent 
spectra we derive a power-law spectral index of 
$\alpha_{\nu} = 0.16 \pm 0.07$ for the entire regime, which is consistent with previous measurements, and fits with the NUV HST spectrum. Although not free of systematics, we consider this the best effort to date to characterize the pulsar SED in this regime, and the result appears to be consistent with synchrotron radiation with little evidence for self-absorption in the NIR.

In addition to the pulsar spectrum, we also highlight a few other aspects of these data. We detected a number of diffuse interstellar bands which have previously not been reported towards the Crab, and also several NIR emission lines, including phosphorus, that to our knowledge have not been previously discussed in the Crab literature. 
All data are now publicly available, and more investigations are encouraged.

\begin{acknowledgements} 
The authors would like to acknowledge 
Claes-Ingvar Bj\"ornsson, Yura Shibanov and Bo Milvang for early discussions on this project.
This work is based on observations collected at the European Organisation for Astronomical Research in the Southern Hemisphere. The Oskar Klein Centre is funded by the Swedish Research Council. 
{Thanks also to the anonymous referee who asked us to look into the broken power-law option.}
\end{acknowledgements}

\bibliographystyle{aa}
\bibliography{crab.bib}

\begin{thebibliography}{54}
\expandafter\ifx\csname natexlab\endcsname\relax\def\natexlab#1{#1}\fi

\bibitem[{{Beskin} \& {Neustroev}(2001)}]{beskin2001}
{Beskin}, G.~M. \& {Neustroev}, V.~V. 2001, \aap, 374, 584

\bibitem[{{Bj{\"o}rnsson} {et~al.}(2010){Bj{\"o}rnsson}, {Sandberg}, \&
  {Sollerman}}]{bjornsson2010}
{Bj{\"o}rnsson}, C.-I., {Sandberg}, A., \& {Sollerman}, J. 2010, \aap, 516, A65

\bibitem[{Boyd(1978)}]{Boyd1978}
Boyd, R.~W. 1978, J. Opt. Soc. Am., 68, 877

\bibitem[{B{\"u}hler \& Blandford(2014)}]{0034-4885-77-6-066901}
B{\"u}hler, R. \& Blandford, R. 2014, Reports on Progress in Physics, 77,
  066901

\bibitem[{{Carrami{\~n}ana} {et~al.}(2000){Carrami{\~n}ana}, {{\v C}ade{\v z}},
  \& {Zwitter}}]{carraminana2000}
{Carrami{\~n}ana}, A., {{\v C}ade{\v z}}, A., \& {Zwitter}, T. 2000, \apj, 542,
  974

\bibitem[{{Crusius-W{\"a}tzel} {et~al.}(2001){Crusius-W{\"a}tzel}, {Kunzl}, \&
  {Lesch}}]{crusiuswätzel2001}
{Crusius-W{\"a}tzel}, A.~R., {Kunzl}, T., \& {Lesch}, H. 2001, \apj, 546, 401

\bibitem[{{Fitzpatrick} \& {Massa}(2007)}]{fitzpatrick2007}
{Fitzpatrick}, E.~L. \& {Massa}, D. 2007, \apj, 663, 320

\bibitem[{{Fordham} {et~al.}(2002){Fordham}, {Vranesevic}, {Carrami{\~n}ana},
  {Michel}, {Much}, {Wehinger}, \& {Wyckoff}}]{fordham2002}
{Fordham}, J.~L.~A., {Vranesevic}, N., {Carrami{\~n}ana}, A., {et~al.} 2002,
  \apj, 581, 485

\bibitem[{{Foreman-Mackey} {et~al.}(2013){Foreman-Mackey}, {Hogg}, {Lang}, \&
  {Goodman}}]{emcee}
{Foreman-Mackey}, D., {Hogg}, D.~W., {Lang}, D., \& {Goodman}, J. 2013, \pasp,
  125, 306

\bibitem[{{Galazutdinov} {et~al.}(2000){Galazutdinov}, {Musaev},
  {Kre{\l}owski}, \& {Walker}}]{galazutdinov2000}
{Galazutdinov}, G.~A., {Musaev}, F.~A., {Kre{\l}owski}, J., \& {Walker},
  G.~A.~H. 2000, \pasp, 112, 648

\bibitem[{{Graham} {et~al.}(1990){Graham}, {Wright}, \&
  {Longmore}}]{graham1990}
{Graham}, J.~R., {Wright}, G.~S., \& {Longmore}, A.~J. 1990, \apj, 352, 172

\bibitem[{{Herbig}(1995)}]{herbig1995}
{Herbig}, G.~H. 1995, \araa, 33, 19

\bibitem[{{Hester}(2008)}]{hester2008}
{Hester}, J.~J. 2008, \araa, 46, 127

\bibitem[{{Hester} {et~al.}(1995){Hester}, {Scowen}, {Sankrit}, {Burrows},
  {Gallagher}, {Holtzman}, {Watson}, {Trauger}, {Ballester}, {Casertano},
  {Clarke}, {Crisp}, {Evans}, {Griffiths}, {Hoessel}, {Krist}, {Lynds},
  {Mould}, {O'Neil}, {Stapelfeldt}, \& {Westphal}}]{hester1995}
{Hester}, J.~J., {Scowen}, P.~A., {Sankrit}, R., {et~al.} 1995, \apj, 448, 240

\bibitem[{{Jones} {et~al.}(2013){Jones}, {Noll}, {Kausch}, {Szyszka}, \&
  {Kimeswenger}}]{Jones2013}
{Jones}, A., {Noll}, S., {Kausch}, W., {Szyszka}, C., \& {Kimeswenger}, S.
  2013, \aap, 560, A91

\bibitem[{{Koo} {et~al.}(2013){Koo}, {Lee}, {Moon}, {Yoon}, \&
  {Raymond}}]{koo2013}
{Koo}, B.-C., {Lee}, Y.-H., {Moon}, D.-S., {Yoon}, S.-C., \& {Raymond}, J.~C.
  2013, Science, 342, 1346

\bibitem[{{Kuiper} {et~al.}(2001){Kuiper}, {Hermsen}, {Cusumano}, {Diehl},
  {Sch{\"o}nfelder}, {Strong}, {Bennett}, \& {McConnell}}]{kuiper2001}
{Kuiper}, L., {Hermsen}, W., {Cusumano}, G., {et~al.} 2001, \aap, 378, 918

\bibitem[{{Loh} {et~al.}(2012){Loh}, {Baldwin}, {Ferland}, {Curtis},
  {Richardson}, {Fabian}, \& {Salom{\'e}}}]{loh2012}
{Loh}, E.~D., {Baldwin}, J.~A., {Ferland}, G.~J., {et~al.} 2012, \mnras, 421,
  789

\bibitem[{{Lundqvist} {et~al.}(1986){Lundqvist}, {Fransson}, \&
  {Chevalier}}]{lundqvist1986}
{Lundqvist}, P., {Fransson}, C., \& {Chevalier}, R.~A. 1986, \aap, 162, L6

\bibitem[{{Lundqvist} \& {Tziamtzis}(2012)}]{lundqvist2012}
{Lundqvist}, P. \& {Tziamtzis}, A. 2012, \mnras, 423, 1571

\bibitem[{{Martin} {et~al.}(1998){Martin}, {Halpern}, \&
  {Schiminovich}}]{martin1998}
{Martin}, C., {Halpern}, J.~P., \& {Schiminovich}, D. 1998, \apjl, 494, L211

\bibitem[{{Massaro} {et~al.}(2006){Massaro}, {Campana}, {Cusumano}, \&
  {Mineo}}]{massaro2006}
{Massaro}, E., {Campana}, R., {Cusumano}, G., \& {Mineo}, T. 2006, \aap, 459,
  859

\bibitem[{{Melatos} {et~al.}(2005){Melatos}, {Scheltus}, {Whiting},
  {Eikenberry}, {Romani}, {Rigaut}, {Spitkovsky}, {Arons}, \&
  {Payne}}]{melatos2005}
{Melatos}, A., {Scheltus}, D., {Whiting}, M.~T., {et~al.} 2005, \apj, 633, 931

\bibitem[{{Mignani} {et~al.}(2012){Mignani}, {De Luca}, {Hummel}, {Zajczyk},
  {Rudak}, {Kanbach}, \& {S{\l}owikowska}}]{mignani2012}
{Mignani}, R.~P., {De Luca}, A., {Hummel}, W., {et~al.} 2012, \aap, 544, A100

\bibitem[{{Mignani} {et~al.}(2007){Mignani}, {Zharikov}, \&
  {Caraveo}}]{mignani2007}
{Mignani}, R.~P., {Zharikov}, S., \& {Caraveo}, P.~A. 2007, \aap, 473, 891

\bibitem[{{Modigliani} {et~al.}(2010){Modigliani}, {Goldoni}, {Royer},
  {Haigron}, {Guglielmi}, {Fran{\c c}ois}, {Horrobin}, {Bristow}, {Vernet},
  {Moehler}, {Kerber}, {Ballester}, {Mason}, \& {Christensen}}]{modigliani2010}
{Modigliani}, A., {Goldoni}, P., {Royer}, F., {et~al.} 2010, in \procspie, Vol.
  7737, Observatory Operations: Strategies, Processes, and Systems III, 773728

\bibitem[{{Moehler} {et~al.}(2014){Moehler}, {Modigliani}, {Freudling},
  {Giammichele}, {Gianninas}, {Gonneau}, {Kausch}, {Lan{\c c}on}, {Noll},
  {Rauch}, \& {Vinther}}]{fluxxshooter}
{Moehler}, S., {Modigliani}, A., {Freudling}, W., {et~al.} 2014, \aap, 568, A9

\bibitem[{{Nasuti} {et~al.}(1996){Nasuti}, {Mignani}, {Caraveo}, \&
  {Bignami}}]{nasuti1996}
{Nasuti}, F.~P., {Mignani}, R., {Caraveo}, P.~A., \& {Bignami}, G.~F. 1996,
  \aap, 314, 849

\bibitem[{{Newville} {et~al.}(2016){Newville}, {Stensitzki}, {Allen}, {Rawlik},
  {Ingargiola}, \& {Nelson}}]{lmfit}
{Newville}, M., {Stensitzki}, T., {Allen}, D.~B., {et~al.} 2016, {Lmfit:
  Non-Linear Least-Square Minimization and Curve-Fitting for Python},
  Astrophysics Source Code Library

\bibitem[{{Noll} {et~al.}(2012){Noll}, {Kausch}, {Barden}, {Jones}, {Szyszka},
  {Kimeswenger}, \& {Vinther}}]{Noll2012}
{Noll}, S., {Kausch}, W., {Barden}, M., {et~al.} 2012, \aap, 543, A92

\bibitem[{{O'Connor} {et~al.}(2005){O'Connor}, {Golden}, \&
  {Shearer}}]{oconnor2005}
{O'Connor}, P., {Golden}, A., \& {Shearer}, A. 2005, \apj, 631, 471

\bibitem[{{Oke}(1969)}]{oke1969}
{Oke}, J.~B. 1969, \apjl, 156, L49

\bibitem[{{Owen} \& {Barlow}(2015)}]{owen2015}
{Owen}, P.~J. \& {Barlow}, M.~J. 2015, \apj, 801, 141

\bibitem[{{Pita} {et~al.}(2014){Pita}, {Goldoni}, {Boisson}, {Lenain}, {Punch},
  {G{\'e}rard}, {Hammer}, {Kaper}, \& {Sol}}]{pita2014}
{Pita}, S., {Goldoni}, P., {Boisson}, C., {et~al.} 2014, \aap, 565, A12

\bibitem[{{Richardson} {et~al.}(2013){Richardson}, {Baldwin}, {Ferland}, {Loh},
  {Kuehn}, {Fabian}, \& {Salom{\'e}}}]{richardson2013}
{Richardson}, C.~T., {Baldwin}, J.~A., {Ferland}, G.~J., {et~al.} 2013, \mnras,
  430, 1257

\bibitem[{{Romani}(2000)}]{romani2000}
{Romani}, R.~W. 2000, in IAU Symposium, Vol. 195, Highly Energetic Physical
  Processes and Mechanisms for Emission from Astrophysical Plasmas, ed.
  P.~C.~H. {Martens}, S.~{Tsuruta}, \& M.~A. {Weber}, 95

\bibitem[{{Romani} {et~al.}(2001){Romani}, {Miller}, {Cabrera}, {Nam}, \&
  {Martinis}}]{romani2001}
{Romani}, R.~W., {Miller}, A.~J., {Cabrera}, B., {Nam}, S.~W., \& {Martinis},
  J.~M. 2001, \apj, 563, 221

\bibitem[{{Rudy} {et~al.}(2015){Rudy}, {Horns}, {DeLuca}, {Kolodziejczak},
  {Tennant}, {Yuan}, {Buehler}, {Arons}, {Blandford}, {Caraveo}, {Costa},
  {Funk}, {Hays}, {Lobanov}, {Max}, {Mayer}, {Mignani}, {O'Dell}, {Romani},
  {Tavani}, \& {Weisskopf}}]{rudy2015}
{Rudy}, A., {Horns}, D., {DeLuca}, A., {et~al.} 2015, \apj, 811, 24

\bibitem[{{Sandberg} \& {Sollerman}(2009)}]{sandberg2009}
{Sandberg}, A. \& {Sollerman}, J. 2009, \aap, 504, 525

\bibitem[{{Sarazin} \& {Roddier}(1990)}]{Sarazin1990}
{Sarazin}, M. \& {Roddier}, F. 1990, \aap, 227, 294

\bibitem[{{Selsing} {et~al.}(2019){Selsing}, {Malesani}, {Goldoni}, {Fynbo},
  {Kr{\"u}hler}, {Antonelli}, {Arabsalmani}, {Bolmer}, {Cano}, {Christensen},
  {Covino}, {D'Avanzo}, {D'Elia}, {De Cia}, {de Ugarte Postigo}, {Flores},
  {Friis}, {Gomboc}, {Greiner}, {Groot}, {Hammer}, {Hartoog}, {Heintz},
  {Hjorth}, {Jakobsson}, {Japelj}, {Kann}, {Kaper}, {Ledoux}, {Leloudas},
  {Levan}, {Maiorano}, {Melandri}, {Milvang-Jensen}, {Palazzi}, {Palmerio},
  {Perley}, {Pian}, {Piranomonte}, {Pugliese}, {S{\'a}nchez-Ram{\'{\i}}rez},
  {Savaglio}, {Schady}, {Schulze}, {Sollerman}, {Sparre}, {Tagliaferri},
  {Tanvir}, {Th{\"o}ne}, {Vergani}, {Vreeswijk}, {Watson}, {Wiersema},
  {Wijers}, {Xu}, \& {Zafar}}]{selsing2019}
{Selsing}, J., {Malesani}, D., {Goldoni}, P., {et~al.} 2019, \aap, 623, A92

\bibitem[{{Serafimovich} {et~al.}(2004){Serafimovich}, {Shibanov}, {Lundqvist},
  \& {Sollerman}}]{serafimovich2004}
{Serafimovich}, N.~I., {Shibanov}, Y.~A., {Lundqvist}, P., \& {Sollerman}, J.
  2004, \aap, 425, 1041

\bibitem[{{Shearer} \& {Golden}(2002)}]{sg2002}
{Shearer}, A. \& {Golden}, A. 2002, in Neutron Stars, Pulsars, and Supernova
  Remnants, ed. W.~{Becker}, H.~{Lesch}, \& J.~{Tr{\"u}mper}, 44

\bibitem[{Smette {et~al.}(2015)Smette, Sana, Noll, Horst, Kausch, Kimeswenger,
  Barden, Szyszka, Jones, Gallenne, Vinther, Ballester, \& Taylor}]{Smette2015}
Smette, A., Sana, H., Noll, S., {et~al.} 2015, A{\&}A, 576, A77

\bibitem[{{Sollerman}(2003)}]{sollerman2003}
{Sollerman}, J. 2003, \aap, 406, 639

\bibitem[{{Sollerman} {et~al.}(2005){Sollerman}, {Cox}, {Mattila},
  {Ehrenfreund}, {Kaper}, {Leibundgut}, \& {Lundqvist}}]{sollerman2005}
{Sollerman}, J., {Cox}, N., {Mattila}, S., {et~al.} 2005, \aap, 429, 559

\bibitem[{{Sollerman} {et~al.}(2000){Sollerman}, {Lundqvist}, {Lindler},
  {Chevalier}, {Fransson}, {Gull}, {Pun}, \& {Sonneborn}}]{sollerman2000}
{Sollerman}, J., {Lundqvist}, P., {Lindler}, D., {et~al.} 2000, \apj, 537, 861

\bibitem[{{Tappe}(2004)}]{Tappe2004}
{Tappe}, A. 2004, PhD thesis, Ph.D dissertation, 2004.~ 138 pages; Sweden:
  Chalmers Tekniska Hogskola (Sweden); 2004.~Publication Number: AAT
  C819472.~DAI-C 66/02, p.~432, Summer 2005

\bibitem[{{Temim} \& {Dwek}(2013)}]{temim2013}
{Temim}, T. \& {Dwek}, E. 2013, \apj, 774, 8

\bibitem[{{Tziamtzis} {et~al.}(2009){Tziamtzis}, {Schirmer}, {Lundqvist}, \&
  {Sollerman}}]{tziamtzis2009b}
{Tziamtzis}, A., {Schirmer}, M., {Lundqvist}, P., \& {Sollerman}, J. 2009,
  \aap, 497, 167

\bibitem[{{Vernet} {et~al.}(2011){Vernet}, {Dekker}, {D'Odorico}, {Kaper},
  {Kjaergaard}, {Hammer}, {Randich}, {Zerbi}, {Groot}, {Hjorth}, {Guinouard},
  {Navarro}, {Adolfse}, {Albers}, {Amans}, {Andersen}, {Andersen}, {Binetruy},
  {Bristow}, {Castillo}, {Chemla}, {Christensen}, {Conconi}, {Conzelmann},
  {Dam}, {de Caprio}, {de Ugarte Postigo}, {Delabre}, {di Marcantonio},
  {Downing}, {Elswijk}, {Finger}, {Fischer}, {Flores}, {Fran{\c c}ois},
  {Goldoni}, {Guglielmi}, {Haigron}, {Hanenburg}, {Hendriks}, {Horrobin},
  {Horville}, {Jessen}, {Kerber}, {Kern}, {Kiekebusch}, {Kleszcz}, {Klougart},
  {Kragt}, {Larsen}, {Lizon}, {Lucuix}, {Mainieri}, {Manuputy}, {Martayan},
  {Mason}, {Mazzoleni}, {Michaelsen}, {Modigliani}, {Moehler}, {M{\o}ller},
  {Norup S{\o}rensen}, {N{\o}rregaard}, {P{\'e}roux}, {Patat}, {Pena}, {Pragt},
  {Reinero}, {Rigal}, {Riva}, {Roelfsema}, {Royer}, {Sacco}, {Santin},
  {Schoenmaker}, {Spano}, {Sweers}, {Ter Horst}, {Tintori}, {Tromp}, {van
  Dael}, {van der Vliet}, {Venema}, {Vidali}, {Vinther}, {Vola}, {Winters},
  {Wistisen}, {Wulterkens}, \& {Zacchei}}]{vernet2011}
{Vernet}, J., {Dekker}, H., {D'Odorico}, S., {et~al.} 2011, \aap, 536, A105

\bibitem[{{Vernet} {et~al.}(2008){Vernet}, {Kerber}, {Saitta}, {Mainieri},
  {D'Odorico}, {Lidman}, {Mason}, {Bohlin}, {Rauch}, {Ivanov}, {Smette},
  {Walsh}, {Fosbury}, {Goldoni}, {Groot}, {Hammer}, {Horrobin}, {Kaper},
  {Kjaergaard-Rasmussen}, {Pallavicini}, \& {Royer}}]{vernet2008}
{Vernet}, J., {Kerber}, F., {Saitta}, F., {et~al.} 2008, in \procspie, Vol.
  7016, Observatory Operations: Strategies, Processes, and Systems II, 70161G

\bibitem[{{Yaron} \& {Gal-Yam}(2012)}]{wiserep}
{Yaron}, O. \& {Gal-Yam}, A. 2012, \pasp, 124, 668

\bibitem[{Zharikov {et~al.}(2007)Zharikov, Mennickent, Shibanov, \&
  Komarova}]{zharikov2007}
Zharikov, S., Mennickent, R.~E., Shibanov, Y., \& Komarova, V. 2007,
  Astrophysics and Space Science, 308, 545

\end{thebibliography}

\clearpage
\begin{deluxetable}{cccc}
\tabletypesize{\footnotesize}
\tablewidth{0pt}
\tablecaption{Diffuse Interstellar Bands towards the Crab pulsar\label{tabledibs}}
\tablehead{
\colhead{DIB\tablenotemark{a}}&
\colhead{Measured Wavelength}&
\colhead{EW}&
\colhead{FWHM}\\
\colhead{(\AA)}&
\colhead{(\AA)}&
\colhead{(\AA)}&
\colhead{(\AA)}\\
}
\startdata
4428    & 4430 & 1.08 & 20 \\
5705.20 & 5707.4 & 0.11 & 3.3 \\
5780.37 & 5782.4 & 0.34 & 2.2 \\
5796.97 & 5799.0 & 0.10 & 1.1 \\
6195.97 & 6198.0 & 0.029 & 0.63 \\
6203.08 & 6205.2 & 0.086 & 1.8 \\
6269.75 & 6271.9 & 0.047 & 1.6 \\
6283.85\tablenotemark{b} & 6286.2 & 0.62 & 4.0 \\
6613.56 & 6615.9 & 0.14 & 1.2 \\
6660.64 & 6658.3 & 0.31 & 11 \\
6993.18 & 6995.4 & 0.037 & 0.62 \\
7223.96 & 7226.4 & 0.15 & 1.3 
\enddata
\tablenotetext{a}{Included are DIBs from \cite{sollerman2005} that were in turn those with $A_{\rm c}\ga0.07$ from the table of \cite{herbig1995}. Detailed rest wavelengths from the \citet{galazutdinov2000} survey. We also added the 4430 DIB.}
\tablenotetext{b}{6379.29 could not be measured since this region was oversubtracted for emission lines, also the 6284 is hampered by this.} 
\end{deluxetable}

\clearpage


\begin{deluxetable}{ccccl}
\tabletypesize{\footnotesize}
\tablewidth{0pt}
\tablecaption{Noticeable emission lines in the NIR\label{tableNIR}}
\tablehead{
\colhead{Ion}&
\colhead{Rest Wavelength}&
\colhead{Velocity}&
\colhead{Dopplershift\tablenotemark{a}}&
\colhead{Comment}\\
\colhead{}&
\colhead{(micron)}&
\colhead{(km~s$^{-1}$)}&
\colhead{(km~s$^{-1}$)}&
\colhead{}
}
\startdata
\ion{H}{I} & 1.282 &  1084 & -164 & P$\beta$, mainly red side \\
\ion{H}{I} & 2.166 &  1038 & -117 & Br$\gamma$, mainly red side \\
\ion{H}{$_2$} & 1.747 & 971  & -14  & potentially  \\
\ion{H}{$_2$} & 2.121 &      &      & not strong, only red side, maybe also 1.957 and 2.033 \\
\ion{He}{I} & 1.012 & 1107 & -141 &  \\  
\ion{He}{I} & 1.083 & 1189 & -74  & several components \\ 
\ion{He}{I} & 1.279 & 968  & -111 & only red reliable \\ 
\ion{He}{I} & 2.059 & 1070 & -167 & red side strong \\ 
$[$\ion{S}{II}$]$ & 1.032 & 731 &  712  & two components of this quadruple? \\
$[$\ion{S}{II}] & 1.037 & 724 &  247  & \\ 
$[$\ion{Fe}{II}] & 1.189 & 1066 & -154  &  \\ 
$[$\ion{Fe}{II}$]$ & 1.257 & 1175 & -172  & strong \\ 
$[$\ion{Fe}{II}$]$ & 1.295 &      &       & weak, only red side \\ 
$[$\ion{Fe}{II}$]$ & 1.321 & 1180 & -190  &  \\
$[$\ion{Fe}{II}$]$ & 1.533 & 1057 & -174  &  red side clear\\
$[$\ion{Fe}{II}$]$ & 1.600 & 1066 & -210  &  red side very clear\\
$[$\ion{Fe}{II}$]$ & 1.644 & 1189 & -170  & strong \\ 
$[$\ion{Fe}{II}$]$ & 1.677 &      &       & only red side \\
$[$\ion{Fe}{II}$]$ & 1.811 & 1015 & -400 & weaker \\
$[$\ion{P}{II}$]$  & 1.189 & 1060 & 493 & Phosphorus, maybe also 1.147 \\ 
\enddata
\tablenotetext{a}{Most lines come both blue and redshifted. The velocities tabulated are  
{$\pm$} 
the velocity offset from centre, and the Dopplershift is a measure of that central wavelength versus the given rest wavelength.
These were measured on the 2D frame of observation number 6. There also seems to be some lines we did not identify, for example at wavelengths 1.38, 1.82, 1.88, 1.95 microns.}
\end{deluxetable}

\clearpage

\appendix

\section{Flux accuracy}

\begin{figure}
   \centering
   \includegraphics[width=9cm]{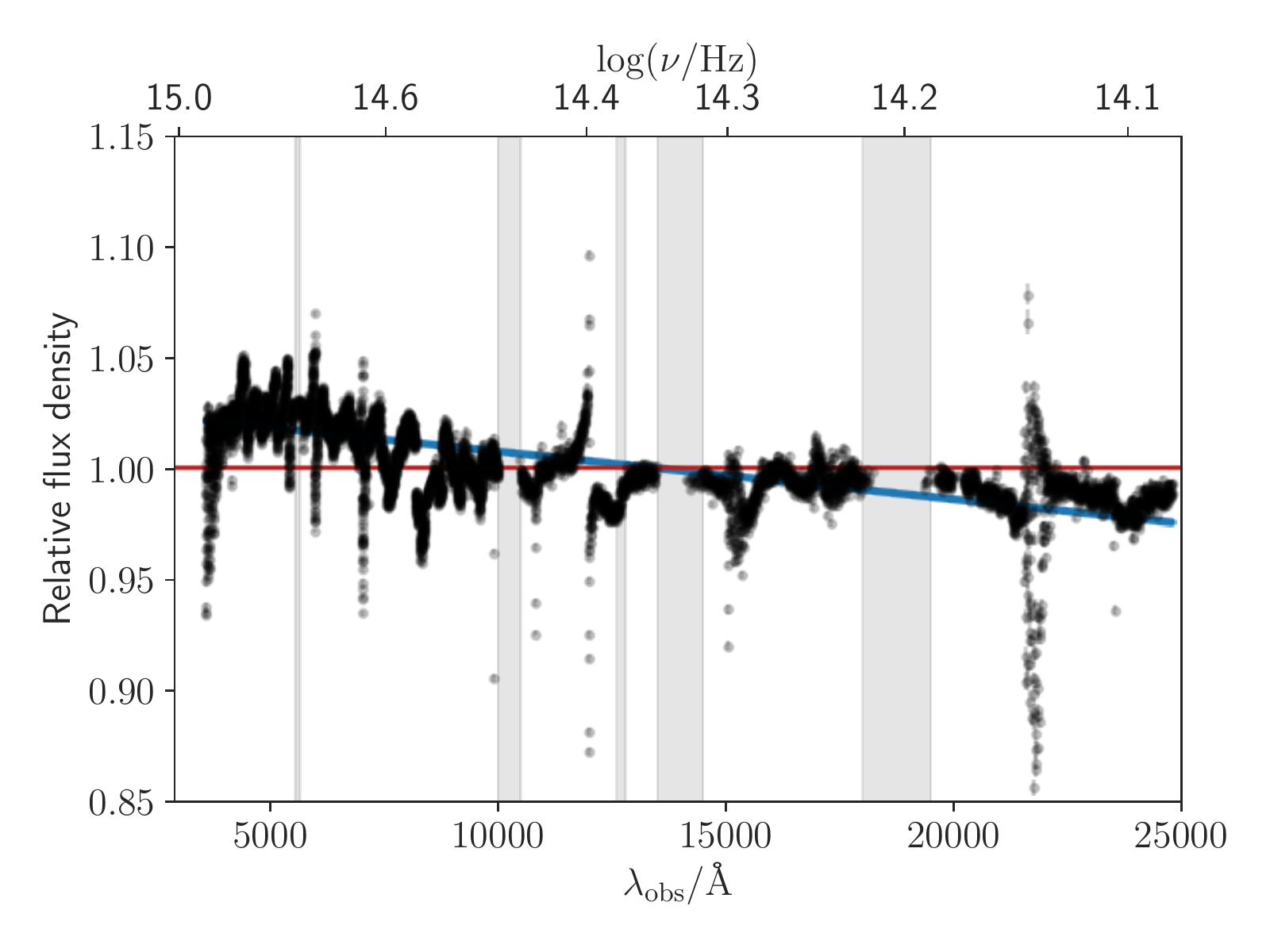}
\caption{
Here we compare the observations we made of the flux-calibrated spectrophotometric standardstar GD71 using the same setup as for the Crab pulsar with a model spectrum of the same star. We have plotted the ratio between these observations, and the blue line is a best fit power-law slope which shows a slight slope which could mean a relative error in flux calibration of $\pm3\%$ over the entire wavelength region of X-shooter.
}
\label{FigApp}
   \end{figure}

As also described in the main text, we observed the spectrophotometric standard star, GD71, using the exact same spectroscopic set-up as for the Crab pulsar observations, including exposure time, readout mode and slit widths (Table~\ref{tab:speclog}). The observations of GD71 were carried out between two consecutive epochs of Crab pulsar X-shooter observations and the comparison between our observed spectra of GD71 and the model spectra presented in \citet{fluxxshooter} should thus represent the relative flux calibration accuracy of the Crab pulsar observations. {We show this in Fig.~\ref{FigApp} which demonstrates an overall relative flux accuracy of $\sim\pm3\%$. We can also see some structure in the ratio, which includes differences between the real stellar spectrum and the model, but also artifacts like echelle order overlaps and residuals from telluric corrections. Again, the overall effect of these are at the percent level, which should be remembered when interpreting wiggles in the pulsar spectrum itself.}

\end{document}